\documentclass{aa} 

\usepackage{graphics}
\usepackage{longtable}
\usepackage{txfonts}
\usepackage{graphicx}
\usepackage{hyperref}
\usepackage{multirow} 
\usepackage{rotating}

\begin{document}

\title{Evidence of an asymmetrical Keplerian disk in the Br$\gamma$ and He~I emission lines around the Be star HD 110432\thanks{Based on observations made with VLTI ESO telescopes at La Silla Paranal Observatory under GTO programme IDs 0.84.C-0062(A), 0.84.C-0062(B), 0.84.C-0062(C), 0.84.C-0062(D)}}

\authorrunning{Ph. Stee et al.}
\titlerunning{HD 110432 circumstellar disk studies with the VLTI/AMBER instrument}

 \author{Ph. Stee\inst{1},  A. Meilland\inst{1,2}, Ph. Bendjoya\inst{1} , F. Millour\inst{1}, M. Smith\inst{3},  A. Spang\inst{1},  G.~Duvert\inst{4}, K.-H.~Hofmann\inst{5}, F.~Massi\inst{6}}

   \offprints{Philippe.Stee@oca.eu}

\institute{Laboratoire Lagrange, UMR 7293 Universit\'e de Nice-Sophia Antipolis (UNS), Observatoire de la C\^ote d�Azur (OCA), Boulevard de l'Observatoire, B.P. 4229 F, 06304 Nice Cedex 4, France.
\and
Physics and Astronomy Department, The University of Western Ontario, London, N6A 3K7, Ontario, Canada.
\and
Catholic University of America, 3700 San Martin Dr., Baltimore, MD 21218 USA.
\and
UJF-Grenoble 1 / CNRS-INSU, Institut de Plan\'etologie et d'Astrophysique de Grenoble (IPAG) UMR 5274, Grenoble, F-38041, France.
\and
Max-Planck-Institut f\"ur Radioastronomie, Auf dem H\"ugel 69, D-53121 Bonn, Germany.
\and
INAF - Osservatorio Astrofisico di Arcetri, Largo E. Fermi, 5, 50125 Firenze, Italy.
}

   \date{Received 29 August 2012 / Accepted 01 December 2012}

   \abstract{{HD 110432 was classified as a ``$\gamma$ Cas X-ray analog" since it 
has similar} peculiar X-ray and optical characteristics, i.e. a hard-thermal X-ray variable emission and an optical spectrum affected by an extensive disk. Lopes de Oliveira et al. (2007) suggest that it might be a Be star harboring an accreting white dwarf or that the X-rays may come from an interaction between the surface of the star and its disk.}
{To investigate the disk around this Be star we used the VLTI/AMBER instrument, which combines high spectral (R=12000) and high spatial ($\theta_{\rm min}$=4\,mas) resolutions.}
{We constrain the geometry and kinematics of its circumstellar disk from the highest spatial resolution ever achieved on this star.}
{We obtain a disk extension in the Br$\gamma$ line of 10.2 D$_{\star}$ and 7.8 D$_{\star}$  in the He\,I line at 2.05 $\mu$m assuming a Gaussian disk model. The disk is clearly following a Keplerian rotation.  We obtained an inclination angle of 55$\degr$, and the star is a nearly critical rotator with V$_{\rm rot}$/V$_{\rm c}$=1.00$\pm$0.2.
This inclination is greater than the value found for $\gamma$\,Cas (about 42$\degr$, Stee et al. 2012), and is consistent with the inference from
optical Fe\,II emission profiles by Smith \& Balona (2006) that the inclination should be more than the $\gamma$\,Cas value. In the near-IR continuum, the disk of HD~110432 is 3 times larger than $\gamma$ Cas's disk.
We have no direct evidence of a companion around HD 110432, but it seems that we have a clear signature for disk inhomogeneities as detected for $\zeta$ Tau.  This asymmetrical disk detection may be interpreted within the one-armed oscillation viscous disk framework.  Another finding is that the disk size in the near-IR is similar to other Be stars with different spectral types and thus may be independent of the stellar parameters, as found for classical Be stars.}{} 


   \maketitle
%

\section{Introduction}
HD 110432 (BZ Cru = HR 4830, B0.5-1 III-IVe; Dachs et al. (1986); Smith \& Balona (2006)) is an interesting object since it exhibits variable hard-thermal X-rays, which is very different from other ``classical" massive stars that are emitting soft X-rays or from non-thermal emission of all known Be/X ray binaries. 
The other well known Be star sharing these properties is $\gamma$ Cas. Smith \& Balona (2006) propose that HD 110432 should be the first 
new member of a select new class of ``$\gamma$ Cas X-ray analogs." These analogs exhibit the same peculiar X-ray and optical characteristics, i.e. Balmer and Fe II double-peaked emission lines with a strong IR excess due to free-free and free-bound emission (Stellebak 1982; Meyer \& Savage 1981), as $\gamma$ Cas itself. HD110432 is not yet known to be in a binary system. However, it may be a member of the cluster NGC\,4609 (Feinstein \& Marraco 1979). In this case its age would be 60\,Myr, and it is a candidate for a blue straggler. In addition, the star is situated near and beyond the Southern Coalsack dark nebula. However, only the UV reddening and extinction are appreciably affected by the presence of this nebula in the foreground.

The X-ray properties of HD\,110432 have been described well by a series of papers by Lopes de Oliveira et al. (2007), Torrej\'on et al. (2012), and Smith et al. (2012). Although the X-ray flaring properties are very comparable, the last two papers demonstrate that the dominant hot thermal plasma component associated with HD\,110432 emission is even hotter than the analogous component of $\gamma$\,Cas.

The origin of this X-Ray emission of these $\gamma$ Cas variables is still a subject of debate. Currently there are two competing scenarios. The first one is based on single Be stars with unusual strong magnetic activity and rapid X-ray correlations with optical and UV activity (see Smith \& Robinson (2003), Henry \& Smith (2012), Smith et al. (2012)).  The second scenario proposes a binary system with a Be star and an accreting degenerate companion, such as a white dwarf in accordance with the evolutionary models of massive binary systems.

To better constrain the physics and the mechanism responsible for this X-ray emission, as well as the link to its circumstellar environment, we observed this Be star with the Very Large Telescope Interferometer (VLTI) combining 1.8m telescopes on baselines up to 128m,  with the AMBER focal instrument that combines high spectral (R=12000) and high spatial ($\theta_{\rm min}$=4\,mas) resolutions (Petrov et al. 2007; Robbe-Dubois et al. 2007) mainly focused on the Br$\gamma$ and He\,I  2.05 $\mu$m emission lines. 
 
In this paper, we would like to address the following points.
 
\begin{itemize}

\item If there is a strong coupling between a putative magnetic field and the circumstellar disk that may lead to part of the observed X-ray emission, is the disk forced to rotate as a solid body by the magnetic field? What is the kinematics within HD110432's circumstellar environment?

\item What is the disk extension in the Br$\gamma$ and He\,I  2.05 $\mu$m emission lines? Is the circumstellar disk  dense and/or large as suggested from the strong and quasi-symmetrical profile of the H$\alpha$ line (EW $\sim$ 60 $\AA$) and the detection of several metallic lines in emission by Lopes de Oliveira et al. (2007) and Torrej\'on et al. (2012)?

\item What is the central Be star's rotational velocity? 
Is the stellar rotation close to critical as already found by Meilland et al. (2012) for eight other Be stars?

\item Since the binarity of HD 110432 may figure in the production of the hard-thermal X-ray emission, either as a source of X-ray emission or previous transfer of angular momentum to the Be star, do we have any evidence of a companion in our interferometric data?  

\end{itemize}

\noindent The paper is organized as follows. In Section 2 we summarize the stellar parameters of HD~110432. In Section 3 we present the VLTI/AMBER observations and the data reduction process. A first analysis using an axi-symmetrical geometrical models is presented in Sects. 4 and 5, and a more advanced modeling of the differential data using a kinematic model is discussed in Section 6. Finally, we summarize our results in Section 7.

\section{The stellar parameters and distance}

To constrain the stellar contribution to the total flux we need to know the effective temperature $T_{\rm eff}$, the radius $R_{\star}$, and the gravity log $g_{\rm eff}$ of the central star. However, these stellar parameters for rapidly rotating Be stars are quite uncertain. Taking gravitational darkening effects into account, Fr\'emat et al. (2005) showed that the apparent $T_{\rm eff}$ is typically thought to be $\sim$ 5 \% to 15\% colder than the values of the non rotating counterparts.

\noindent In the case of HD~110432, they found an apparent $T_{\rm eff}$ of 20324$\pm$344K and $T_{\rm eff}$ of 24070$\pm$603 for its parent non rotating counterpart (\textit{pnrc}). Using the same method, Zorec et al. (2005) derived an average T$_{\rm eff}$ on the stellar surface of 22510\,K. Knowing these uncertainties on the definition of the effective temperature, we adopted $T_{\rm eff}$=22000$\pm$2000K in this paper.

\noindent On the other hand, Fr\'emat et al. find an apparent g$_{\rm eff}$ of 3.638$\pm$0.042 and 3.950$\pm$0.074 for the \textit{pnrc}. As this parameter does not strongly affect the overall SED and the determination of the stellar fluxes in the infrared, we simply assume g$_{\rm eff}$=3.9$\pm$0.1.

\noindent The stellar radius is also strongly affected by the stellar rotation. Zorec et al. (2005) then estimate the stellar mass as $M/M_{\odot}=9.6$ and the stellar age $t/t_{\rm MS}=0.61$  ($t_{\rm MS}$ is the time the rotating star spend in the main sequence), which mean that the corresponding `spherical' radius of the parent non rotating object is $R_\star$=5.3$R_\odot$, which leads to the critical radius at the equator of 7.7$R_\odot$. We note that the mean apparent radius will depend on the object inclination angle. However, here we decide to adopt a value of $R_\star$=6.5$\pm$1.2R$_\odot$. Using the distance derived from van Leeuven (2007) Hipparcos parallaxes, i.e. d=373$\pm$pc, we infer a photosphere angular size of 0.16 mas. The parameters are summarized in Table~\ref{stellar_param}. 

\noindent Using this set of stellar parameters and photometric measurements from the SIMBAD \footnote{available at \url{http://simbad.u-strasbg.fr/simbad/}} database, we tried to determine the relative flux of the circumstellar environment within the VLTI/AMBER spectral domain, i.e. the H and K bands. To determine the stellar flux in these bands we used Kurucz models reddened using the law of extinction from Cardelli et al. (1989). However, we found that the influence of the reddening in the near-infrared fluxes is negligible compared to the uncertainties on the stellar parameters and distance.

Finally we found that the emission in the H and K bands are both dominated by the circumstellar environment with a relative circumstellar flux (F$_{\rm disk}$) of 71$\pm$7$\%$ and  79$\pm$5$\%$, respectively. 

\begin{table}[!t]
\centering \begin{tabular}{ccccc}
\hline
Parameter			&T$_{\rm eff}$	&	log g$_{\rm eff}$	& R$_\star$			& d	\\
(unit)				&(K)						&			-							&	(R$_\odot$)		&	(pc)\\
\hline\hline
Value					& 22000					&		3.9							&		6.5					&	373\\
Uncertainty 	&	$\pm$2000			&		$\pm$0.1				&	$\pm$1.2			& $\pm$41	\\
\hline
\end{tabular}
\caption{Stellar parameters and distance of HD\,110432  adopted in this paper.\label{stellar_param}}
\end{table} 

\begin{table*}[!htb]
\centering \begin{tabular}{ccccccccc}
\hline \multicolumn{3}{c}{Observation}   & \multicolumn{2}{c}{Projected baseline}& Mode & DIT & Seeing & Calibrators\\
\hline      Date& Time          & Triplet  & L(m)          & P.A. ($^o$)				&       & (s) & (")    &    (HD)\\
\hline
\hline
24/01/2010    & 07:32     & A0-K0-G1 & 78.7/81.1/127.9& -163.2/-89.5/-125.7  & HR-K-F-2.17     &12   & 0.69 & 113752\\
26/01/2010    & 07:33     & A0-K0-G1 & 78.5/81.8/127.8& -161.7/-87.4/-123.7  & HR-K-F-2.17     &12   & 0.60 & 113752\\
10/02/2010    & 07:33     & A0-K0-G1 & 76.1/86.0/126.2& -151.5/-73.6/-109.7  & HR-K-F-2.17     &12   & 0.66 & 113752\\
10/02/2010    & 08:29     & A0-K0-G1 & 73.2/88.3/123.5& -143.1/-62.4/-98.2   & HR-K-F-2.06     &12   & 0.84 & 113752\\
17/03/2010    & 07:30     & D0-H0-G1 & 68.2/58.5/53.5 & -31.8/98.9/24.1      & LR-HK-F         &0.05 & 0.36 & 113752\\
17/03/2010    & 07:41     & D0-H0-G1 & 68.4/58.0/53.1 & -29.8/101.5/25.5     & LR-HK-F         &0.05 & 0.50 & 113752\\
17/03/2010    & 08:15     & D0-H0-G1 & 69.0/56.1/51.5 & -23.1/109.6/30.0     & LR-HK-F         &0.05 & 0.57 & 113752\\
19/03/2010    & 05:23     & A0-K0-G1 & 75.6/86.5/125.8& -149.8/-71.4/-107.4  & LR-HK-F         &0.05 & 0.41 & 113752\\
19/03/2010    & 06:00     & A0-K0-G1 & 73.6/88.1/123.9& -144.0/-63.7/-99.5   & LR-HK-F         &0.05 & 0.37 & 113752\\
19/03/2010    & 06:34     & A0-K0-G1 & 71.3/89.2/121.5& -138.7/-56.4/-92.0   & LR-HK-F         &0.05 & 1.30 & 113752\\
19/03/2010    & 07:08     & A0-K0-G1 & 68.6/89.9/118.5& -133.4/-49.2/-84.4   & LR-HK-F         &0.05 & 0.90 & 113752\\
\hline
\end{tabular}
\caption{Observations log of HD 110432}
\end{table*}

\section{Observations and data reduction process}
\begin{figure}[!t]
\centering   
\includegraphics[width=0.43\textwidth]{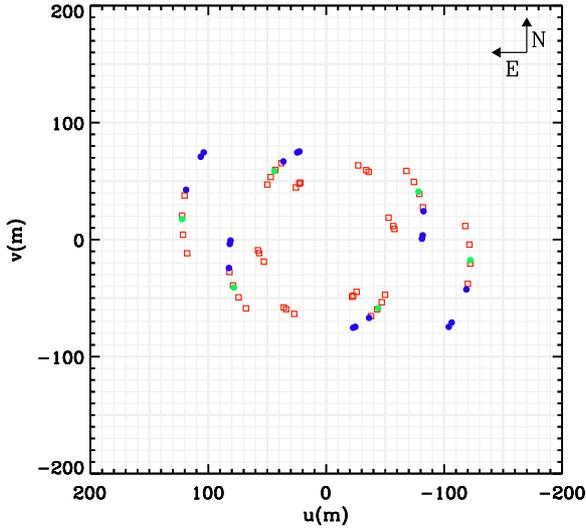}
\caption{(u,v) plane coverage for our VLTI/AMBER observations of HD 110432.  LR observations are plotted as red squares, HR data as blue circles for Br$\gamma$, and green ones for the 2.06 $\mu$m He\,I line.}
\label{uv}
\end{figure}

HD~110432, was observed in January, February and March 2010 with the  A0-K0-G1 and D0-H0-G1 VLTI/AMBER triplets of the 1.8m Auxiliary Telescopes array, ranging from 51 to 124m baselines (See Fig.~\ref{uv}). We used both low resolution (LR=30) and high resolution (HR=12000) measurements in the H (1.54-1.87$\mu$m) and K (1.94-2.37$\mu$m) bands with the help of the FINITO fringe tracker. With an apparent diameter of $0.83 \pm 0.01$ mas in the CHARM2 catalog from Richichi et al. (2005), HD 113752 was used to calibrate the visibility measurements (see Table 2). More details about the AMBER instrument can be found in Petrov et al. (2007).

Data were reduced using the VLTI/AMBER data reduction software, i.e., \texttt{amdlib v3.0.3b1} (see Tatulli et al. 2007 and Chelli et al. 2009 for detailed information on the AMBER data reduction). We selected individual exposures with the standard selection criteria (Millour et al. 2007). We rejected  the 80$\%$ of frames with the lowest S/N. For observations in LR mode we also rejected the frames with a piston larger than 10$\mu$m, as well as frames with a flux ratio between the beams higher than three.

\begin{figure*}[t]
\centering
 \includegraphics[width=0.9\textwidth]{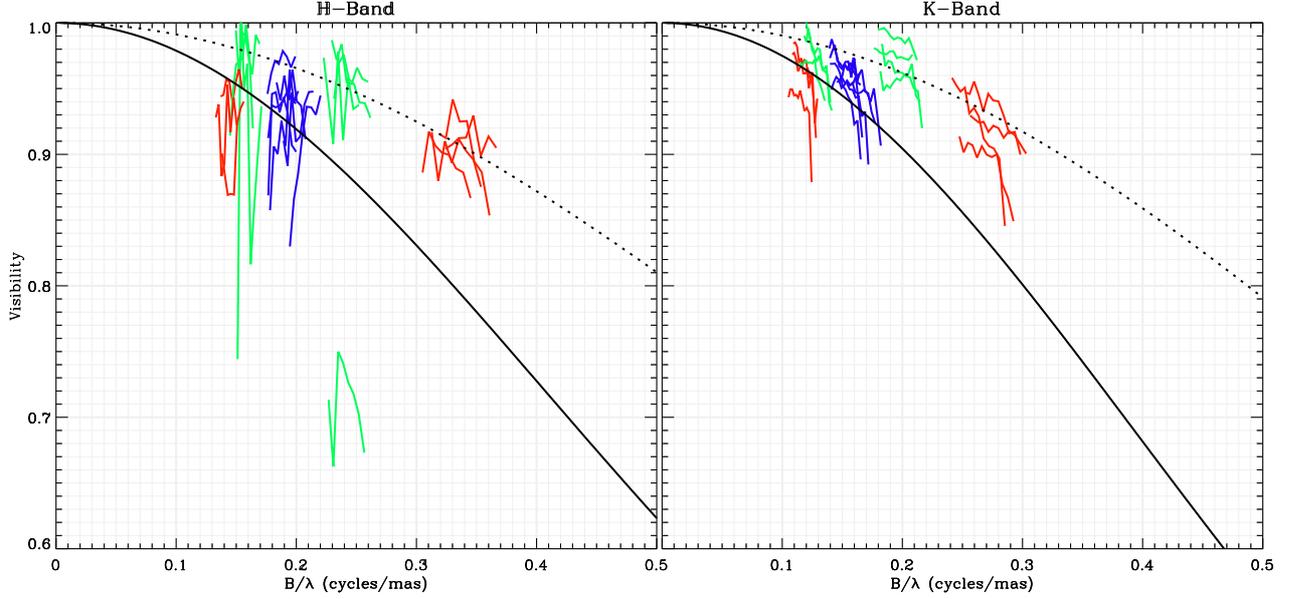} 
\caption{\textbf{Left.} H band data from the VLTI/AMBER instrument for different baselines (blue, red and green lines) with a LITpro model over-ploted along the minor (dotted line) and major (plain line) disk axis.  \textbf{Right.} Same kind of data but for the K band.}\label{image_Litpro}
\label{data}
\end{figure*}

The interferometric observables (visibility, differential phase, and closure phase) were then averaged and calibrated. For this last step we used scripts described in Millour et al. (2007) that are now part of the standard amdlib package. The calibration process includes an estimation of the calibrators' size and their uncertainties from various catalogs, a determination of the transfer functions and their evolution during the whole night, and a computation of the calibrated visibilities and phases. The final errors on the measurements include uncertainties on the calibrators' diameter, the atmosphere transfer function fluctuations, and intrinsic errors on the measurements (Fig.~\ref{data}).

\section{Modeling the H and K band continuum}
\subsection{The LITpro software}

To model the continuum visibility modulus and closure phases obtained from our VLTI/AMBER observations we used the \texttt{LITpro}\footnote{LITpro software available at http://www.jmmc.fr/litpro}  model-fitting software for optical/infrared interferometric data developed by the Jean-Marie Mariotti Center (JMMC) to analyze our data (Tallon et al. 2008). It is based on the Levenberg-Marquardt algorithm, which allows a fit to converge to the closest $\chi^2$ local minimum from a set of initial values of the model parameters. It also includes tools for facilitating the search for the global minimum. LITpro calculates an error on the fitted parameters based on the $\chi^2$ value at the minimum. It uses data error estimates based on the OI FITS format, which does not include error correlation estimates. Therefore, in some cases, LITpro can provide underestimated errors on the parameters.

\subsection{Extension of HD110432 in the continuum }
Even if the spatial resolution of the VLTI/AMBER instrument is about 4 mas, it is not necessary to fully resolve the target, i.e. to reach the first zero of the visibility function to estimate its size. It is always possible to use a partial resolution of the interferometer to estimate the size of the target with a model that fits the data and thus determines the angular size of targets smaller than 4 mas. See for instance the review of the interferometric observations of rapidly rotating stars by van Belle (2012).  Thus, to estimate the disk extension in the near infrared, we model the central star as a uniform disk  + an extended elliptical Gaussian distribution. This model then has four free parameters: the FWHM of the major axis of the Gaussian distribution ($\theta_{\rm disk}$), the relative environment continuum flux (F$_{\rm disk}$), the flattening ratio of the major to the minor Gaussian distribution axis (f), and the position angle of the major axis of the disk measured eastward from the north (P.A).

\noindent Since we have VLTI/AMBER data in two different spectral domains, namely the H band between 1.54-1.87 $\mu$m and the K band within 1.94-2.37 $\mu$m, we tried to constrain both data sets separately. HD~110432 is a variable Be star, and the data used to estimate the circumstellar contribution to the total near-IR continuum flux is not contemporaneous  to our interferometric measurements. Nevertheless, using the values of F$_{\rm disk}$ obtained in section 2, i.e. 71$\pm$7$\%$ and 79$\pm$5$\%$ in the H and K bands respectively, we can infer the disk dimension in the continuum.

\begin{figure}[!h]
\centering
 \includegraphics[width=0.45\textwidth]{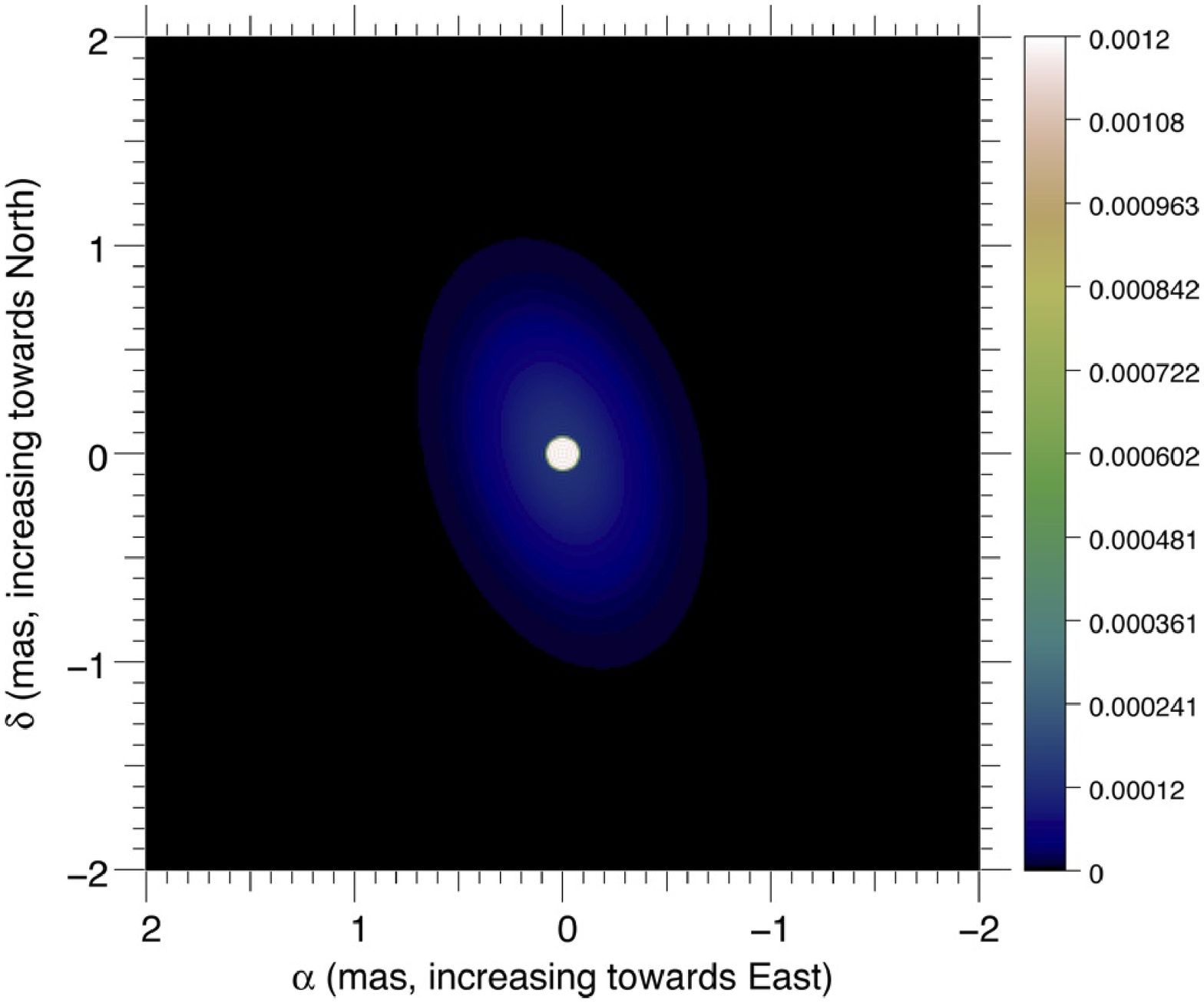} 
 \caption{LITpro model of HD 110432 Cas disk in the K band assuming a uniform central star with a surrounding Gaussian disk}		
\label{image_K}
\end{figure}

\noindent The best LITpro models and the corresponding extensions are given in Table~\ref{Lit_Pro_results}. We obtain a major Gaussian distribution axis of 0.92 mas, a flattening ratio of 1.56 in the H band that, assuming a flat disk, corresponds to 50 $\pm$ 5$\degr$ with a reduced $\chi^{2}_r$ of 4.4 and a P.A. of 22$\pm$ 5$\degr$. In the K band the disk appears to have similar properties: an angular size of 0.95 mas, a flattening of 1.62 corresponding to an inclination inclination angle of  52 $\pm$ 5$\degr$ and a compatible P.A of 17 $\pm$ 5$\degr$.  These extensions are similar to the disk size in the near-IR continuum, i.e. 0.82$\pm$0.08 mas, obtained for $\gamma$ Cas by Stee et al. (2012). Nevertheless, for  $\gamma$ Cas it corresponds to 1.9 D$_{\star}$ and thus HD~110432 exhibits a disk size three times larger in the near-IR continuum than $\gamma$ Cas. An intensity map in the K band is plotted in Fig.~\ref{image_K}. The P.A. of $\sim$ 20 $\degr$ is neither perpendicular nor parallel to the polarization measurement of 81$\degr$ by Yudin (2001). The error bars on these disk measurements are mainly dominated by the uncertainties on the disk flux contribution in the H and K bands.

\begin{table}[!b]
\caption{\label{Lit_Pro_results}Parameters obtained with \texttt{LITpro} for HD 110432 assuming a 0.31 mas central star (see section 2).}
{\centering \begin{tabular}{lclc|cl}
\hline
 Parameters  									&        H band	& K band \\	
 											&	1.54-1.87$\mu$m   						&	1.94-2.37$\mu$m  \\
\hline \hline
$\theta_{\rm disk}$	(mas)						&       0.92 $\pm$0.05  					& 0.95 $\pm$0.07 \\
$\theta_{\rm disk}$	(D$_{\star}$)					&	5.7		&	5.9 \\
f					  						&       1.56 $\pm$0.05		& 1.62 $\pm$0.03  \\
P.A. 											&	      22 $\pm$5$\degr$								& 17 $\pm$5$\degr$  \\
Inclination (i)											& 50 $\degr$	& 52 $\degr$ \\										
F$_{\rm disk} (\%)$											& 			71 $\pm$7								& 79 $\pm$5	\\
\hline
$\chi^{2}_r$ 				 								&       4.4									&	3.4   \\
\hline
\end{tabular}\par}
\tablefoot{$\theta_{\rm disk}$ is the FWHM of the major axis of the Gaussian distribution, F$_{\rm disk}$ the relative environment continuum flux, $\rm f$ the flattening ratio of the major to the minor Gaussian distribution axis, and P.A. the position angle of the major axis of the disk measured eastward from the north. The estimated inclination angle (i) is given assuming a very thin disk.}
\end{table}

\begin{figure*}[!th]
\centering   
\includegraphics[width=0.93\textwidth]{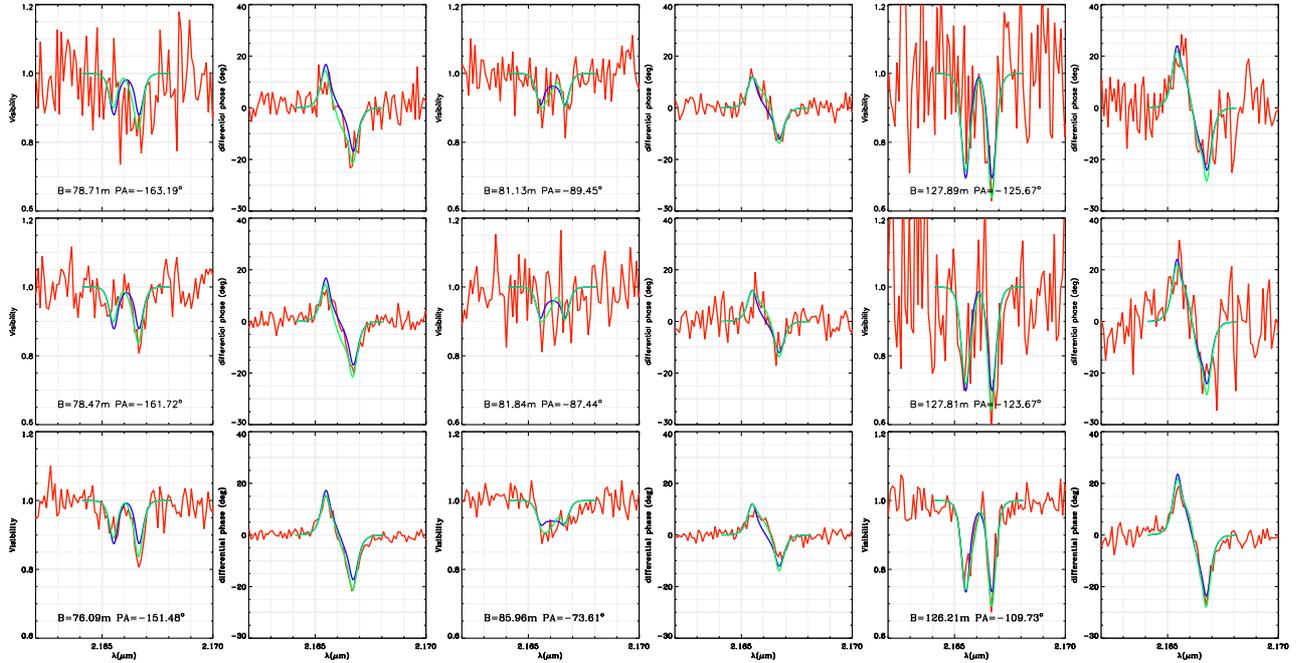}
\vspace{0.1cm}
\caption{HD 110432 differential visibility and phase within the Br$\gamma$ line from our four VLTI/AMBER HR measurements (red line). Each row corresponds to one VLTI/AMBER measurement (3 different baselines). The visibility and phase of the best-fit kinematics model is overplotted in green.}
\label{BrG_vis}
\end{figure*}
\begin{figure*}[!th]
\centering   
\includegraphics[width=0.99\textwidth]{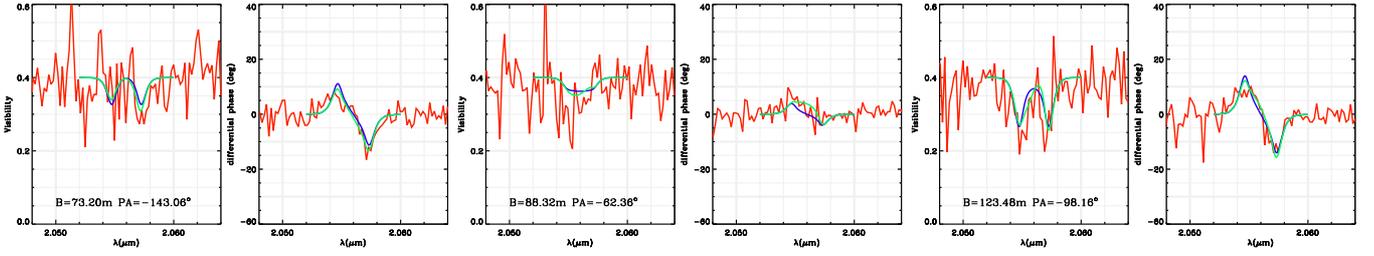}
\vspace{0.1cm}
\caption{HD 110432 differential visibility and phase within the He\,I line from our four VLTI/AMBER HR measurements (red line). The differential visibility is first plotted with the corresponding phase for 3 different baselines with B=73.20m, PA=-143.06$\degr$;   B=88.32m, PA=-62.36$\degr$; and B=123.48m, PA=-98.16$\degr$. The visibility and phase of the best-fit kinematics model is overplotted in green.}
\label{HeI_vis}
\end{figure*}

\section{Modeling the disk in Br$\gamma$ and HeI emission lines}

\subsection{The kinematic model}

To quantitatively constrain the velocity fields in the circumstellar environment of HD110432, we developed a simple two-dimensional kinematic model of a rotating and/or expanding equatorial disk. This model has already been used to successfully model classical Be stars (see Delaa et al. 2011; Meilland et al. 2011; Meilland et al. 2012)  and is described in detail in Delaa et al. (2011). The model geometry is completely ad-hoc: the star is modeled as a uniform disk and the envelope emission in the emission line as an elliptical Gaussian distribution with a given FWHM and a flattening due to a projection effect of the geometrically thin equatorial disk, i.e., $f = 1/cos(i)$, where i is the star-disk system's inclination angle. 

The emission maps are then combined with a two-dimensional projected velocity map of a geometrically thin expanding and/or rotating equatorial disk. For each spectral channel in the line, an iso-velocity map projected along the line of sight is then calculated and multiplied by the whole emission map in the line. 

The model parameters can be classified into three categories:  

\begin{enumerate}
\item The stellar parameters: stellar radius ($R_\star$), distance ($d$), inclination angle ($i$), and disk major-axis position angle ($PA$).
\item The kinematic parameters: rotational velocity ($V_{\rm rot}$) as measured from the stellar surface, expansion velocity at the photosphere ($V_0$), terminal velocity ($V_\infty$), and exponents of the expansion ($\gamma$) and rotation ($\beta$) velocity laws. 
\item The disk emission line parameters: disk FWHM in the line ($a_{\rm l}$) and line equivalent width (EW). 
\end{enumerate}
The star distance is taken from van Leeuwen (2007), and $R_\star$ is derived from the fit of the SED. The nine other parameters are free. 

If the disk is directly connected to the stellar surface, the rotational velocity (V$_{\rm rot}$) should be equal to the stellar rotational velocity. However, in some cases, V$_{\rm rot}$ may exceed the stellar velocity if the star is not critically rotating and some additional momentum is transferred to the circumstellar matter. Finally, we considered in our modeling that V$_{\rm rot}$ is a free parameter with a higher maximum value equal to the critical velocity (V$_{\rm c}$).

\subsection{Model fitting and results}

We have computed several hundred models to constrain the parameters, determined the uncertainties, and tried to detect any degeneracy or linked parameters. Owing to the large number of free-parameters, an automatic model-fitting method would have resulted in computing millions of models. Moreover, we clearly know each parameter's effect on the visibility and phase variations (see Meilland et al. 2012). Consequently, we decided to perform the fit manually. We could exclude models with significant expansion velocity of more than a few km\,s$^{-1}$. Consequently, we decided to set the expansion velocities to zero. We then tried to constrain the seven remaining parameters ($i$, $PA$, $V_{\rm rot}$, $\beta$, $a_{\rm c}$, $a_{\rm l}$, and $EW$). To reduce the number of computed models, we started with a qualitative estimation of the parameters from our interferometric data (especially for $PA$, $i$, a$_c$, a$_l$, and $EW_l$) and explored the parameter space with decreasing steps to converge to a $\chi^2$ minimum. To check for other minima, we also explored the full range of possible parameters space but with larger steps. 

The parameter values for the best-fit model are presented in Table~\ref{model_params}. The corresponding differential visibilities and phases are overplotted in Figs.~\ref{BrG_vis} and \ref{HeI_vis}. For this model, we obtained a reduced $\chi^2$ of 2.0. The overall morphology and amplitude of the lines' profile and  differential visibilities and phases are roughly fitted. Nevertheless,  such a simple axisymmetric model cannot reproduce the asymmetries of the spectro-interferometric data, and this issue will be discussed in more detail in Section 6.3.

\begin{table}[!ht]
\caption{Parameters values for the best-fit axisymetric-kinematic model.\label{model_params}}
\centering \begin{tabular}{ccc}
\hline
Param.				& Value				&Remarks\\
\hline\hline
\multicolumn{3}{c}{\textbf{Global geometric parameters}}\\
$R_\star$				&	6.5						 R$_\odot$				& \\
$d$							&	373						pc							& From von Leeuween (2007)\\
$i$	  					& 55 $\pm$ 5$\degr$						 & \\
$PA$						& 45 $\pm$ 10$\degr$						& Inconsistent with Polar.\\
\hline
\multicolumn{3}{c}{\textbf{Global kinematic parameters}}\\
$V_\mathrm{rot}$ 				&	450	$\pm$50		\,km\,s$^{-1}$ & See discussion\\
$\beta$						      &	0.5	$\pm$0.05									 & Keplerian rotation\\
\hline
\multicolumn{3}{c}{\textbf{Br$\gamma$ disk geometry}}\\
$a_\mathrm{Br\gamma}$ 	&	10.2$\pm$0.5				D$_\star$				& = $1.64 \pm 0.08 $mas\\
$EW_\mathrm{Br\gamma}$	&	10$\pm$0.5			$\AA$						&\\
\hline
\multicolumn{3}{c}{\textbf{He\,{\sc i} disk geometry}}\\
$a_\mathrm{He{\sc I}}$ 			&	7.8$\pm$1.0 D$_\star$ 		    	&= $1.25 \pm 0.16$  mas\\
$EW_\mathrm{He{\sc I}}$		  &	8.$\pm$1.0 $\AA$						&\\
\hline\hline
\end{tabular}
\end{table}

\section{Discussion}\label{discussion}

\subsection{The central star}
We obtain the same inclination angle i.e. i=55 $\pm$ 5$\degr$, similar to the one determined from the continuum fit in the previous section and a P.A. of 45 $\pm$ 10$\degr$, larger than the P.A obtained in the H and K bands ($\sim$ 20$\pm$ 5$\degr$). This is more than the value found for $\gamma$\,Cas (about 42$\degr$, Stee et al. 2012), and is consistent with the inference from optical Fe\,II emission profiles by Smith \& Balona (2006) that the inclination should be greater than the $\gamma$\,Cas value. It is also consistent with the value of  $68^o\pm5^o$ obtained from the evolutionary tracks by Ekstr\"om et al. (2012).

The disk P.A. is neither perpendicular nor parallel to the polarization measurement of 81$\degr$ by Yudin (2001).  We obtain a rotational velocity of $450\pm50$ km\,s$^{-1}$. Assuming that V$_{\rm rot}$, the velocity at the base of the disk is the stellar rotational velocity, we obtain a V\,sin\,i of 368$\pm$43 km.s$^{-1}$. This is higher than the value determined by Slettebak (1982), i.e. $V\!\sin i\simeq300$ km\,s$^{-1}$, and Ballereau et al. (1995), i.e. $V\!\sin i\simeq318 $km\,s$^{-1}$,  but compatible with $V\!\sin i=400\pm30$ km\,s$^{-1}$ determined by  Chauville et al. (2001) from model fitting of the HeI 4471 and MgII 4481 lines, duly corrected for the veiling effect and considering wavelength-dependent limb darkening inside the spectral lines.
 
As already noted in Sect. 2, Zorec et al. (2005) estimate that the stellar mass is $M/M_{\odot}=9.6$. Consequently, this gives a critical velocity  for the star of $V_{\rm c}=487\pm32$ km\,s$^{-1}$.  With this value, an inclination angle of 55$\pm$5$^o$, and Chauville et al.(2005) V\,sin\,i of 400$\pm$30km\,s$^{-1}$, we obtain V$_{\rm rot}$/V$_{\rm c}$=1.00$\pm$0.2. Consequently we can conclude that HD\,110432 is rotating very close to its breakup velocity, like almost all interferometrically observed Be stars.

\subsection{The disk kinematics and extension}
\noindent The disk is rotating following a Keplerian rotation, as is now clearly established for Be stars (Meilland et al.  2007; Meilland et al. 2012). Since this is the first interferometric observation of HD\,110432, it is difficult to compare the geometry obtained. However, we can compare disk radii in the
Br$\gamma$ with the measurements obtained in a survey of eight Be stars by Meilland et al. (2012). Using the same equipment and reduction algorithms we found that in Br$\gamma$ flux at least the disk size of HD\,110432 is larger to those of $\alpha$ Col (5.5 $\pm$ 0.3 R$_{\star}$) and $\alpha$ Ara (5.8 
$\pm$ 0.5 R$_{\star}$). Following Meilland et al. (2012), we note that $\alpha$ Col is a B7IV star with a T$_{eff}$ of 12963 $\pm$ 203 K and $\alpha$ 
Ara a B3IV star with a T$_{eff}$ of 18044 $\pm$ 312 K. Since both are considerably cooler than HD 110432, conditions other than only the stellar effective temperature do influence the disk size. 

We recall that the total emitted flux in the Br$\gamma$ line has two components. One of them is due to the photospheric absorption of the star that underlies the circumstellar disk or envelope. The other component corresponds to the emission produced in the circumstellar environment. 

For this simple model, we can write the expression of the modulus of the visibility in the Br$\gamma$ line as

\begin{equation}
V_{\rm Br\gamma} = \frac{V_{\rm\star Br\gamma}F_{\rm\star Br\gamma}+V_{\rm env Br\gamma}F_{\rm env Br\gamma}}{F_{\rm tot}}  ,
\label{eq2}
\end{equation}

\noindent where $V_{\rm\star Br\gamma}$ and $F_{\rm\star Br\gamma}$ represent the visibility and the flux of the photospheric Br$\gamma$ absorption respectively, while $V_{\rm env Br\gamma}$ and $F_{\rm env Br\gamma}$ are the visibility and the flux emitted by the envelope in the Br$\gamma$ line. The $F_{\rm tot}$  flux is simply

\begin{equation}
F_{\rm tot} = F_{\rm\star Br\gamma} + F_{\rm env Br\gamma}.
\label{eq3}
\end{equation}

\noindent The quantity of interest in our analysis of the emitting region 
($V_{\rm env Br\gamma}$) can then be written by using Eq.\ref{eq2} and Eq.\ref{eq3}:

\begin{equation}
V_{\rm env Br\gamma} = \frac{V_{\rm Br\gamma}-V_{\rm\star Br\gamma}\frac{F_{\rm\star Br\gamma}}{F_{\rm tot}}}{1-\frac{F_{\rm\star Br\gamma}}{F_{\rm tot}}}.
\label{eq4}
\end{equation}

\noindent Thus, changing the effective temperature will change the ratio $\frac{F_{\rm\star Br\gamma}}{F_{\rm tot}}$ in a nontrivial way since it may increase $F_{\rm\star Br\gamma}$ but also change the value of $F_{\rm env Br\gamma}$. This is because modifying T$_{eff}$ will also change the source function of the disk emitting region. The same argument applies to the formation of He\,I line and nearby continuum emission.

\noindent From this paper and the cited work, the disk size does not seem to be correlated with the stellar spectral type. We also measured the disk size within the He\,I emission line at 2.05 $\mu$m, which seems to be smaller compared to the Br$\gamma$ line, i.e. 7.8 vs 10.2 D$_{\star}$. To our knowledge, the only other Be star with a disk measurement in the He\,I 2.05 $\mu$m emission line was the binary Be star $\delta$ Sco with a disk extension of  4.5 $\pm$ 0.5 D$_{\star}$ (Meilland et al. 2011), thus smaller than our measurement by a factor 1.7. In the Br$\gamma$, Meilland et al. (2011) find a FWHM of 5.5 $\pm$ 1D$_\star$, again 1.8 times smaller than the one we measured for HD\'110432. The difference may originate from the youth of $\delta$ Scorpii disk, which started to form in 2000 and which was still growing in 2010 according to the same authors. Note that Meilland et al. (2011) were able to measure the disk of $\delta$ Sco in the Br$\gamma$ line out to the inferred disk truncation radius, i.e. $\sim$ 2.24 mas or 4.2 R$_{\star}$ at periastron. But this truncation radius is impossible to determine for HD\,110432 since the mass and the separation of the system components (assuming the star is a binary) are unknown.

\noindent  In a recent paper, Touhami et al. (2011) have estimated the near-infrared continuum emission from the circumstellar gas disk of Be stars using a radiative transfer code for a parametrized version of the viscous decretion disk model. They were able to predict the half-maximum emission radius along the major axis of the projected disk in the H and K bands. For HD 110432 they obtain a color excess $E^{\star}(V^{\star}-18 \mu m)$ of 2.5 which corresponds to a disk density of $\sim$ 3.1 10$^{-11}$ g cm$^{-3}$  from their Fig.~1. Using their Fig.~2 and Table 1, it translates into a disk HWHM radius of $\sim$1.41-1.58 R$_{\star}$ so smaller than our $\sim$ 5.8 R$_{\star}$. To obtain a similar disk size, it seems from their Table, that the disk density is certainly higher, i.e. $\sim$ 1.0~10$^{-11}$ g cm$^{-3}$. On the other hand, many effects may explain this discrepancy:

\begin{itemize}
\item A larger disk inclination and a density of $\sim$ 8.1~10$^{-11}$ g cm$^{-3}$. The estimated radius by Touhami et al. (2011) becomes 3.13  R$_{\star}$.
\item The contamination of the interferometric visibilities from the incoherent light of a possible companion close enough to influence our signal at the time of the VLTI observations. 
In that case, a correction of the visibilities from this contamination is necessary before solving for the angular diameter of the disk. That would considerably lower the
fitting value of angular size. Although Mason et al. (1997) did not detect a companion through speckle, neither do we, and the system is suspected to be a high mass X-ray binary.
\item Variability: our interferometric measurements were taken at the beginning of 2010, whereas the AKARI IR excesses originated before 2010.
\item  The value for the density exponent adopted in the model (n=3) by Touhami et al. (2011) . This assumption could be argued to justify the higher disk size. In fact, the
IR excess and the disk size are highly sensitive to this parameter.
\item The HWHM used in their model is the image disk radius, which could be slightly different from the disk radius obtained from a Gaussian elliptical disk.
\item A higher density than what the model predicts, which may be a clue to a new outburst episode. This could be verified by looking for variabilities in H$\alpha$ EW, but unfortunately we were not able to  find any H$\alpha$ spectra before 2010.
\end{itemize}

\subsection{Disk inhomogeneities}

\begin{figure*}[!ht]
\centering   
\includegraphics[width= 0.40\textwidth]{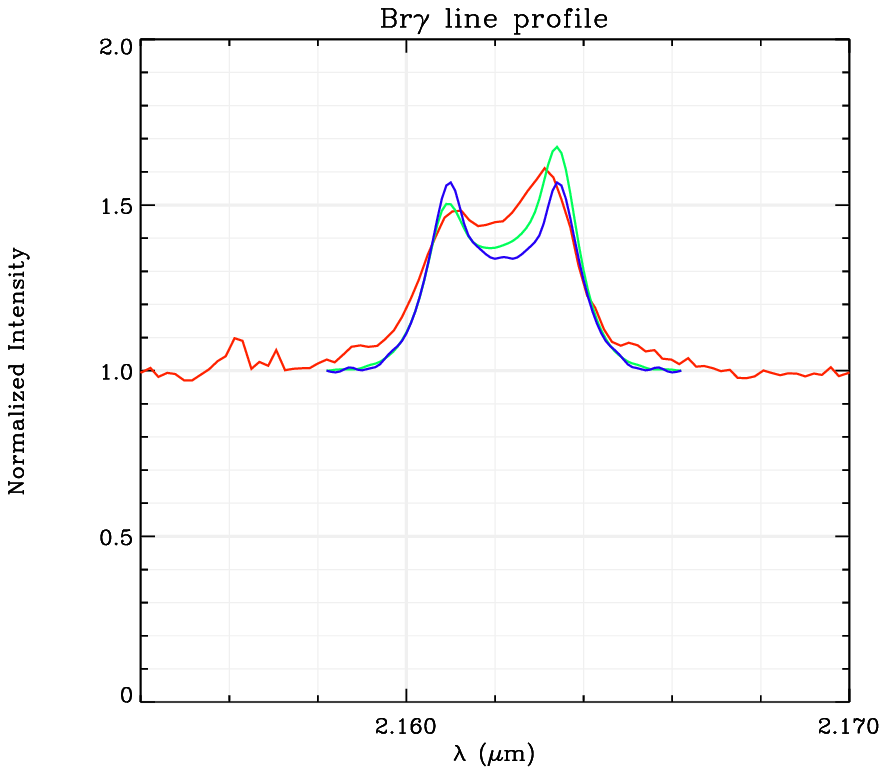}\hspace{1.0cm}
\includegraphics[width=0.40\textwidth]{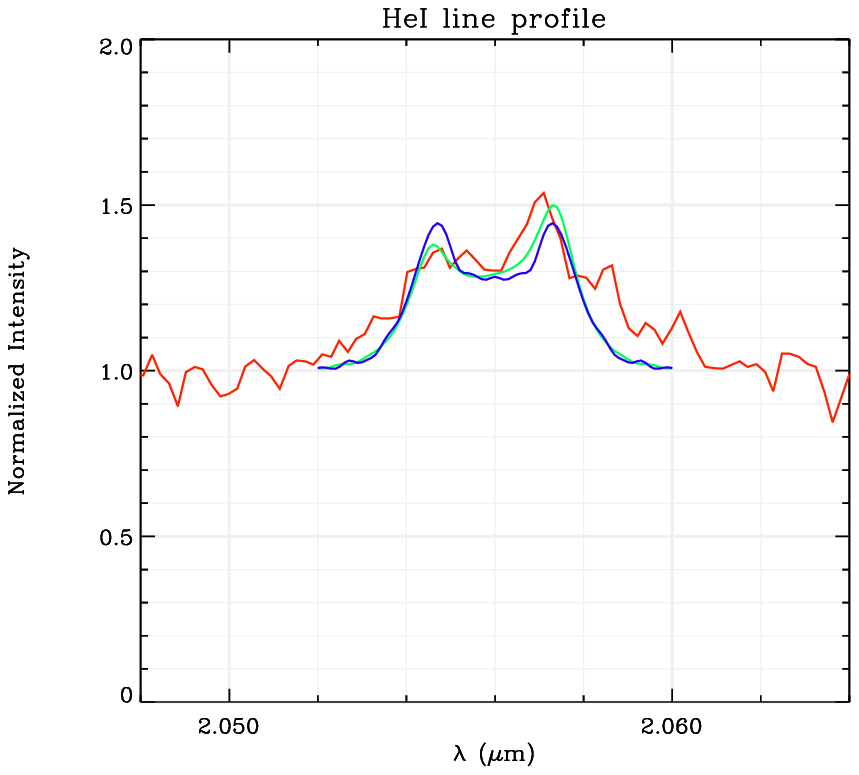}
\vspace{0.5cm}
\caption{Br$\gamma$ (Right) and {\bf He\,I} lines from our VLTI/AMBER observations (red line) with our symmetrical line profile fit (blue line) and our asymmetrical model (green line).}
\label{lines}
\end{figure*}

\noindent We see from Fig~\ref{BrG_vis} and  Fig~\ref{HeI_vis} that the differential phases for these two lines are, first, represented well by our very simple kinematical model, and second that they exhibit the typical ``S" shape of a rotating disk. The amplitude of this ``S" shape is $\sim 10-15 \degr$, which is very similar to the differential phases obtained by Carciofi et al. (2009) for the Be star $\zeta$ Tau, and for $\delta$ Sco by Meilland et al. (2011). Our differential phases exhibit a very asymmetrical ``S" shape with a smaller amplitude in the blue part of the ``S" curve with respect to the central line wavelength. This was also the case for $\zeta$ Tau. This was clear evidence for a one-armed spiral structure in the $\zeta$ Tau disk, which may also be the case for HD 110432. Moreover, our visibilities across the Br$\gamma$ line are asymmetrical, with a red wing of the visibility systematically smaller than the blue wing. The line profiles are also clearly asymmetrical, with Br$\gamma$ and He\,I line profiles showing a V/R\,$< 1$ even if the S/N and spectral resolution for He\,I are not high enough to clearly believe this line asymmetry {\bf (See Fig.~\ref{lines})}. Our simple symmetrical ``toy" model is not able to reproduce the V/R of the observed profiles as seen from Fig.~\ref{lines}. \\

\noindent Thus, HD\,110432 circumstellar disk is globally rotating,  with a larger volume and brighter region responsible for the red part of the line, i.e. flowing away from us, and a smaller (more compact), fainter, or more absorbed region in the blue part of the line, i.e. rotating in our direction. These emitting regions are also responsible for the Br$\gamma$ line profiles with V/R\,$< 1$. Moreover, since the differential phase is asymmetrical, it means that the photocenter of the emitting regions is asymmetric with respect to the central star (or the rotational axis). 

\noindent To account for these asymmetries, we tested two different types of asymmetric models. In the first one the line emission is modeled by an asymmetric gaussian disk :

\begin{equation}
I_l(r,\theta)=e^{-\frac{r^2}{2\sigma^2(\theta)}}~~~\rm{with}~~\sigma(\theta)=\sigma_0(1+A\,cos(\theta-\theta_0))
\end{equation}

In this model the Gaussian width, given by $\sigma$, depends on the azimuth $\theta$. The reference for the azimuth ($\theta$=0$^o$) is the projected polar axis of the star. The parameters $A$ and $\theta_0$ set the amplitude and orientation of the asymmetry, respectively.

In the second type of asymmetric model, we try to roughly reproduce the one-arm oscillation model presented in Okazaki (1997). Thus, we consider an antisymmetric spiral-like intensity distribution given by

\begin{equation}
I_l(r,\theta)=e^{(-\frac{r^2}{2\sigma^2})\,}(1+A\,cos(\theta-\theta_0-\pi r/T))
\end{equation}

\noindent where $A$ is the amplitude of the spiral perturbation of the Gaussian distribution, and $T$ the spiral period.\\

\begin{figure}[!t]
\centering   
\includegraphics[width=0.50 \textwidth]{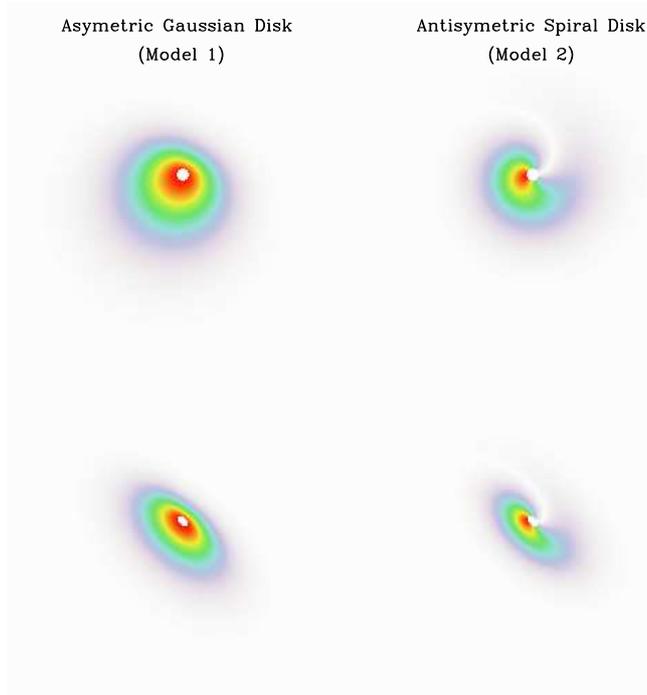}
\caption{Examples of pole-on (Upper figures) and projected onto the sky (Lower figures) intensity distributions for the two asymmetric classes of models tested in this paper. The central white circle represents the position of the central star.}
\label{asym_distrib}
\end{figure}

\noindent Examples of both intensity distributions are shown in Figure~\ref{asym_distrib}. As for the axi-symmetric model, the intensity distribution is then rotated on the skyplane and flattened, taking the inclination angle of the object into account. The use of these models significantly improve the fit of the data. The $\chi^2_r$ obtained for the best-fit asymmetric Gaussian disk and the antisymmetric spiral are 1.4 and 1.6, respectively, compared to 2 for the symmetrical model (see section 5). For both types of models, the best-fit model was obtained without modifying the stellar and kinematic parameters from Table~\ref{model_params} and with only minor changes in terms of the FWHM of the intensity distribution and/or equivalent width of the emission line. The corresponding parameters values for these models, namely $A$, $\theta_0$, and $T$,  are presented in Table~\ref{param-asym}.\\

\noindent The two additional parameters for the asymmetric Gaussian disk, i.e. A and $\theta_0$ are well constrained. The asymmetry is quite large, i.e. the largest disk extension is about 1.7 times the smallest one and the over intensity is orientated close to the projected minor axis, and mainly to the south-east. In the case of the antisymmetric spiral model, we could not obtain asymmetric visibility variations for a period of the spiral arms of less than T=20 R$_\star$. Actually, for shorter periods, the asymmetries introduced in the model by the spiral averaged themselves out, considering that the disk FWHM in the emission lines are on the order of 4-5R$_\star$. Nevertheless, for longer periods, we were  still  able to constrain the two other parameters for this model. As for the asymmetric Gaussian disk model, we obtain an asymmetry that is roughly oriented in the south-east.

\noindent Finally, the better agreement of the asymmetric Gaussian disk with the interferometric data is mostly because this model is able to produce stronger asymmetries in the visibility with almost symmetric S-shaped phases. Moreover, this model also reproduces the visibility and phases observed along the polar orientation better. Nevertheless, the quantity and quality of our data are not sufficient to fully constrain the intensity distribution in the emission lines. 

\noindent Regarding the possibility of a companion, we have no evidence of binarity from our interferometric measurements, neither from a modulation of the visibility modulus as a function of time or spatial frequency, nor from a phase modulation of the fringes. The question of an unseen companion orbiting HD 110432 becomes an issue when studying the origin of its hard X-ray flux, if indeed the X-rays are formed somehow from infall in the form of  gas stream or blobs originating in the Be star.

\begin{table}[!ht]
\caption{Parameters values for the best-fit {\bf asymmetric} kinematic model.\label{param-asym}}
\centering \begin{tabular}{ccc}
\hline
~~~~Parameters~~~~~~				& ~~~~~~Model 1~~~~~~				&~~~~~~Model 2~~~~\\
\hline\hline
\hline
\multicolumn{3}{c}{\textbf{Br$\gamma$ disk geometry}}\\
$a_\mathrm{Br\gamma}$ 	&	9.8$\pm$0.2D$_\star$			         & 10.2$\pm$0.2D$_\star$\\
$EW_\mathrm{Br\gamma}$	&	9.5$\pm$0.5$\AA$						        & 10$\pm$0.5$\AA$	\\
\hline
\multicolumn{3}{c}{\textbf{He\,{\sc i} disk geometry}}\\
$a_\mathrm{He{\sc I}}$ 			&	7.4$\pm$0.5 D$_\star$ 		    	&	7.6$\pm$0.5 D$_\star$\\
$EW_\mathrm{He{\sc I}}$		  &	8.$\pm$1.0 $\AA$						    &8.$\pm$1.0 $\AA$\\
\hline\hline
\multicolumn{3}{c}{\textbf{Asymmetry}}\\
A&	0.25$\pm$0.05	&	0.55$\pm$0.1\\
$\theta_0$& 165$\pm5^o$	&225$\pm10^o$\\
T&	-	&	$\geq$20	\\
\hline\hline

\end{tabular}
\end{table}

\section{Conclusion}

Using standard long baseline optical interferometric techniques, we have reported on the size and orientation of the disk of the first ``$\gamma$\,Cas X-ray analog" system other
than the prototype $\gamma$\,Cas itself. We obtained a disk FWHM of 10.1 D$_{\star} $ in the Br$\gamma$ and 7.7 D$_{\star}$ in the He I lines. \\

\noindent We have clearly detected an asymmetry in the disk in both the Br$\gamma$ and He~I lines which is to our knowledge the first time that an asymmetry is detected on such a small spatial scale within the He~I emission line. As a whole, the viscous disk model seems to be a possible scheme to match the inhomogeneities detected in HD 110432 disk since it produces non symmetrical, spectrally resolved visibilities and phases, as measured by Carciofi et al. (2009). Nevertheless, the pursuit of possible disk inhomogeneities requires a more dedicated set of observations, as done by Carciofi et al. (2009) for  $\zeta$ Tau.\\

\noindent The disk is in Keplerian rotation and is in contact with the central star, as for $\gamma$ Cas. If HD 110432 has a magnetosphere, it must be very small, i.e.  smaller than the size of classical T-Tauri magnetospheres that lie between 3 R$_{\star}$ and 7 R$_{\star}$  (Getman et al. 2008) and probably smaller than 1 R$_{\star}$, in order to have undetectable effects on our interferometric measurements.

\noindent The disk's major-axis position angle (PA) of HD 110432 is 45$\degr$ from the kinematics fit or 20$\degr$ from the continuum disk emission fit, but in both cases  is neither aligned nor perpendicular to the polarization measurements. Finally the disk size seems to be independent of the stellar parameters, as already found for classical Be stars by Meilland et al. (2012).\\

\noindent We concluded that HD 110432 is a near critical rotator characterized  by V$_{\rm rot}$/V$_{\rm c}$=1.00$\pm$0.2.\\


\begin{acknowledgements}
This work has benefited from funding from the French Centre National de la Recherche Scientifique (CNRS) through the Institut National des Sciences de l'Univers (INSU) and its Programmes Nationaux (ASHRA, PNPS).This research made use of SIMBAD and VIZIER databases, operated at the CDS, Strasbourg, France. We thank Yamina Touhami for an interesting discussion about the continuum disk size and the referee for his extensive and extremely helpful criticisms of the original manuscript. 
\end{acknowledgements}

\end{document}